\documentclass[twoside]{dis07}
\usepackage[latin1]{inputenc}
\usepackage[dvips]{graphicx,epsfig,color}
\usepackage{wrapfig,rotating}
\usepackage{amssymb,amsmath,array,amsbsy}

\def\vec#1{\boldsymbol{#1}}
\begin{document}
\title{Charmonium singlets, open charm and exotic hadrons}

%***********************************************************************
% AUTHORS INFORMATION AREA
%***********************************************************************
\author{Jean-Marc Richard%$^1$ and Second Author$^2$
%
% Optional short acknowledgment: remove next line if non-needed
%\thanks{This is an optional funding source acknowledgment.}
%
% DO NOT MODIFY THE FOLLOWING '\vspace' ARGUMENT
\vspace{.3cm}\\
%% Addresses and institutions (remove "1- " in case of a single institution)
LPSC, CNRS-IN2P3-Ing\'enierie--UJF--INPG \\
53, avenue des Martyrs, 38026 Grenoble, France%
}
%***********************************************************************
% END OF AUTHORS INFORMATION AREA
%***********************************************************************
\maketitle

\begin{abstract}
Caution is suggested on the comparison of the spin-singlet charmonium P-state with the centre of gravity of triplet states, when the mass splitting is of the order of a few MeV. The physics of new hidden-charm  states $X$ and $Y$ is briefly reviewed. Perspectives for producing double-charm  baryons and double-charm exotic mesons are discussed.
\end{abstract}

\section{Charmonium singlets}
The charmonium singlet states have resisted a firm identification  for many years \cite{Martin:2003rr}, but, now the $\eta_c$, $\eta_c'$ and $h_c$ are well identified. The mass of the latter, as given by the CLEOc collaboration is \cite{Rubin:2005px} $m(h_c)=3524.4\pm 0.6\pm0.4\;\mathrm{MeV}/c^2$, where the first uncertainty  comes from statistics and the second one from estimated systematic errors. For the previous attempts and other measurements, see, e.g., \cite{Andreotti:2005vu,DIS2007}. 
%In particular a slightly higher mass is found in the reaction $\mathrm{\bar{p}p}\to \eta_c\gamma$ \cite{}.

It is perfectly legitimate to \emph{define} the hyperfine splitting by $\delta=(\chi_0+3\chi_1+5\chi_2)/9 -h_c$. This leads to the experimental value  \cite{Rubin:2005px}
%\begin{equation}\label{HF1P}
$\delta= 1.0 \pm 0.6\pm 0.4\;\mathrm{MeV}/c^2$~.
%\end{equation}
%
But a superficial reading of $\delta$ could be misleading, as it suggests a very small or even vanishing effect of spin--spin forces in the 1P multiplet, while it is arguably larger, and positive, of the order of a few MeV.

In the potential models, if the spin-orbit, $V_{LS} \vec{L}.\vec{S}$ and tensor, $V_T S_{12}$ terms are treated in perturbation, the masses of triplet P-states with $J=0$, $1$ and $2$ are shifted  by 
$\{-2,-1,1\}\langle V_{LS}\rangle$ and $\{-4,2,-2/5\}\langle V_T\rangle$, respectively,
and it is readily seen that the contributions of $\langle V_{LS}\rangle$ and $\langle V_T\rangle$ disappear in the $(2J+1)$-weighted centre of gravity. Now the $\chi_2-\chi_0$ splitting is of the order of 150\;MeV, suggesting that at the level of 1\;MeV accuracy, the calculation of the spin splittings should be pushed beyond first order.  As the spin operators $\vec{L}.\vec{S}$ and $S_{12}=3\vec{\sigma}_1.\hat{r}\,\vec{\sigma}_2.\hat{r}-\vec{\sigma}_1.\vec{\sigma}_2$ enter the Hamiltonian linearly,  the ground state of the spin-triplet P-state is a concave function of the values of  these operators. This means that in any potential model where  the components $V_{LS}$ and $V_T$ are suitably regularised and inserted non-pertubatively into the wave equation,  the genuine spin-orbit- and tensor-free triplet state is \emph{above} the naive centre of  gravity. As an example, if one adopts the potential of Ref.~\cite{Vijande:2004he}, which is rather conventional in the heavy-quark sector (the light-quark one is more speculative, with Goldstone-boson exchanges), the difference is about 3\;MeV, which locates the experimental $h_c$ about 4\;MeV above this improved triplet benchmark.

Unfortunately, the convexity effect depends rather sensitively on the details of the regularisation of the spin-orbit and tensor terms,  and many other effects have to be taken into account, for instance, the P--F orbital mixing, which pushes down  the $^3\mathrm{P}_2$ state.
Phenomenological potentials could also include further spin operators that are not necessary in the simplest non-relativistic reduction of one-gluon exchange, or in the Thomas precession, and, for spin triplet,  cannot be reduced to  spin--orbit and tensor and thus do not average out to zero in the naive centre of gravity. An example is the ``quadratic spin--orbit'' operator used in nuclear physics to 
describe the nucleon--nucleon potential.

On the theoretical side,  the zero-range property of the spin-spin interaction, $V_{SS}\propto \delta^{(3)}(\vec{r})$ only holds in the simplest non-relativistic reduction \`a la Breit--Fermi of the  Coulomb term due to one-gluon-exchange. A range of about the inverse quark mass would be reasonable, and would give a non-vanishing matrix element for P-states. Higher-order terms in perturbative QCD have been discussed in the literature, see, e.g.,\cite{Brambilla:2004wf,Recksiegel:2003fm} and Refs.~there.
The spin--spin potential has also been estimated non-perturbatively with the lattice techniques 
\cite{Brambilla:2004wf}, exhibiting a range that is short but finite. In some other lattice studies 
\cite{Brambilla:2004wf}, the states $\chi_J$ and $h_c$ are calculated directly (not \`a la Born--Oppenheimer through  a potential), each with a specific operator adapted to its quantum numbers, as in QCD sum rules \cite{Narison:2004}.

The interpretation of the mass of the $\eta'_c$ also reveals the limits of simple potentials. The splitting 
$m(\psi')-m(\eta_c')\simeq 48\;\mathrm{MeV}$,
is appreciably smaller than the predicted one in most $(c\bar{c})$ constituent models tuned to reproduce $m(\mathrm{J}/psi)-m(\eta_c)=117\;$MeV. A likely explanation is that for this 2S multiplet lying very close to the $\mathrm{D}\overline{\mathrm{D}}$ threshold, the coupling to higher Fock configurations is enhanced.  The $\psi'$ is pushed down by the very close  $\mathrm{D}\overline{\mathrm{D}}$ threshold, while $\eta'_c$ is less affected, since only influenced by the higher lying $\mathrm{D}^*\overline{\mathrm{D}} + \mathrm{c.c.}$  and  $\mathrm{D}^*\overline{\mathrm{D}}{}^*$ thresholds. This reduces the effect of the quark--antiquark spin--spin forces \cite{Martin:1982nw}.

Hopefully, the $\eta_b$ will be found shortly. Its mass can be estimated with sophisticated techniques.  If one accounts for the $m^{-2}$ factor in front of the spin-spin interaction, and the squeezing of wave-functions when the constituent masses increase, which for a logarithmic potential gives a factor $m^{3/2}$ for the squared wave function at the origin, one gets $m(\Upsilon)-m(\eta_b)$ in ratio $(m_c/m_b)^{1/2}$ to its charm analogue, i.e., about 68\;MeV.
\section{Single and double charm hadrons}
A few years ago, several new results came in the meson sector with single charm, in particular the $\mathrm{D}_{s,J}$ states, and this stimulated an intense activity.  More recently, several new baryons have been found, and nowadays, the family of charmed baryons include many states  \cite{DIS2007}.

A key question in baryon spectroscopy is to find evidence for the three-body structure, i.e., states in which both $\vec{x}=\vec{r}_2-\vec{r}_1$ and $\vec{y}\propto 2 \vec{r}_3-\vec{r}_1-\vec{r}_2$ degrees of freedom are excited. In the harmonic-oscillator of light quarks, with flavour SU(3) symmery, this corresponds to the $20^+$ multiplet, with an antisymmetric orbital  wave-function $\psi(\vec{x},\vec{y})\propto \vec{x}\times\vec{y}\exp[-a(x^2+y^2)/2]$ that couples to an antisymmetric spin--isospin wave function and an antisymmetric colour wave function. The lack of firm experimental candidate is perhaps due to the small  cross-sections in pion- or photo-production experiments, which favour states having better overlap with the quark wave function of the target nucleon. Another picture is proposed by diquark models, in which these states do not exist, if the diquark is in its ground state.

Perhaps the first baryon with excitation in both Jacobi variables will be found in the charm sector: this state is expected to be rather narrow and to have preferentially at least one orbital excitation in its decay products.

Among ordinary hadrons, the $(QQq)$ baryons with two heavy quarks are particularly interesting, as they combine  the adiabatic motion of two heavy quarks, as in charmonium, and the relativistic motion of a light quark around a coloured source, as in $\mathrm{D}$ mesons. 

The ground state has interesting weak-decay properties.There are huge differences among the lifetimes of $\mathrm{D}$, $\mathrm{D}_s$ mesons and single-charm baryons. The hierarchy of the lifetimes is well understood in terms of $W$-exchange, or interference effects, but the differences are usually larger than estimated in calculations. In the case of hadrons with two heavy quarks, binding effects also play a role.

SELEX has serious candidates for the ground states, and  more fragile evidence for the isospin partner, spin or orbital excitations \cite{Engelfried:2007at}.
However, other experiments were not able to find any double-charm baryons, in particular in $e^+e^-$ \cite{Aubert:2006qw}. This is a little surprising, because meanwhile the B-factories found an excess (vs.~simple QCD expectations) of double charm-pair production, $e^+e^-\to (c\bar{c})(c\bar{c})$, leading to beautiful $(c\bar{c})$ spectra recoiling against the J/$\psi$. One would naively expect that if a $(cc\bar{c}\bar{c})$ primary system is easily produced, it sometimes rearranges into  a doubly-charm diquark and a conjugate antidiquark.

When the first studies of double-charm baryons were carried out, their experimental study was out of reach. Now, with the $\mathrm{B}_c$ well measured, and the first indication for $\Xi_{cc}$, the sector of two heavy flavours seems ready for detailed spectroscopy, and one could already envisage one step beyond, i.e., triple charm.
The spectrum of $\Omega_{ccc}$ was called \cite{Bj} ``the ultimate goal of baryon spectroscopy'', the true analogue of charmonium  for baryons. Here the three-quark dynamics can be tested in the static  limit and confronted with theory. For instance, the level order is expected to be similar to that of charmonium, with the first excitation having a parity opposite to that of the ground-state. Remember that for light baryons, the Roper resonance, with the same positive parity as the ground-state nucleon, comes slightly lower than the first orbital excitations, and this cannot be accommodated in simple quark models.
\section{Crypto-exotic and exotic hadrons}
Several intriguing states have been identified in the hidden-charm sector,  that are hardly compatible with genuine $(c\bar{c})$ states. The experimental situation concerning the  $X(3872)$, the various 
$X$'s near 3940, the $Y(4260)$, etc., is reviewed in several contributions to this conference \cite{DIS2007}.

The most popular explanation of $X(3872)$ is that of a $\mathrm{D}\overline{\mathrm{D}}{}^*+\mathrm{c.c.}$ molecule, see, e.g., \cite{Swanson:2006st} and Refs.~there on the pioneering works by Voloshin et al., T\"ornqvist, Glashow et al.,  Ericson and Karl, Manohar et al., Braaten et al., etc. Nuclear forces acting between charmed mesons generate a nuclear potential which is weaker that the proton--neutron spin-triplet interaction, but being experienced by heavier particles, it gives comparable spectral properties, at the edge between binding and non-binding.
Interesting developments have been proposed, in particular bound sates of two or several charmed or doubly-charmed baryons \cite{Froemel:2004ea}. Also, as the $\mathrm{D}$ and the $\overline{\mathrm{D}}{}^*$ are not strictly bound in this approach, but slightly above their threshold, one could envisage the Borromean binding of three or more  heavy mesons.

There are, however, some caution in order. As stressed by Suzuki \cite{Suzuki:2005ha}, due to the $\mathrm{D}^*-\mathrm{D}$ mass difference, the Yukawa potential in $\mathrm{D}\overline{\mathrm{D}}{}^*+\mathrm{c.c.}$ is non local, and this might weaken its efficiency. Also, the miracle in nuclear physics is the presence of a hard core, which prevents the nucleons from collapse and reinforces the role of long-range dynamics. There is no such hard core in $\mathrm{D}\overline{\mathrm{D}}{}^*+\mathrm{c.c.}$, and one should account for the direct interaction between the quarks of $\mathrm{D}$ and these of $\overline{\mathrm{D}}{}^*$. This leads us to the alternative four-quark models.

Among these models, there is the diquark--antidiquark picture, as developed in particular by Maiani  et al.\ \cite{Maiani:2005pe}, giving an unified picture of several new states. Notice that the diquark is an effective cluster, an approximation valid only in a given environment.  If taken too seriously, some of the diquark  models  of $X$ and $Y$, with a relatively low mass for the diquarks, could lead to predict the existence of stable triple-charm dibaryons, such as $(cccsss)$, below the $\Omega\Omega_{ccc}$ threshold.

The four-quark dynamics, and its application to $X(3872)$, is also discussed by Lipkin and H\o gaasen  et al.\ \cite{Hogaasen:2005jv}, among others. The chromomagnetic interaction, with a realistic treatment of flavour-symmetry breaking gives a simple explanation of the mass and decay properties of $X(3872)$ \cite{Hogaasen:2005jv}.

Now, a lesson from atomic physics, is that the best place for stable four-body states is not $(M^+,m^+,M^-,m^-)$, which is slightly stable for $M=m$, but loses stability for $M/m\simeq 2.2$ \cite{Armour:2005}. However, the configurations $(M^+,M^+,m^-,m^- )$ are more stable that these with  equal masses \cite{Armour:2005}. The crucial rule is that the Coulomb interaction remains unchanged when the masses evolve from electron to muon or heavier constituents. In QCD, we have the same property, called \emph{flavour independence}, for the spin-independent interaction. This is why states of the type $(QQ\bar{q}\bar{q})$ are predicted to exist \cite{Ader:1981db}, at least in the limit of large $Q/q$ mass ratio. Their production and identification could be carried out in the experiments searching for the double-charm baryons.

%\section{Outlook}
A good surprise of recent high-energy experiments has been the ability of performing productively  in hadron physics, and even to clarify the results claimed by dedicated low-energy experiments. Another good surprise is the ability to produce fragile and complex structures, such as antideuterium \cite{DIS2007}, bound only by 2\;MeV. It is reasonable to anticipate significant progress on heavy hadrons, in particular exotic multiquarks, from the future LHC experiments, provided a small fraction of the analysis is devoted to this physics.

I thank for organisers of  this beautiful DIS Conference, K. Seth of discussions there, and M.\ Asghar
for comments on the manuscript, and J.\  Vijande for correspondence.

%\section*{Acknowledgments}
%Discussions at this Conference with K.~Seth on charmonium singlets, and earlier with E.~Eichen, A.~Martin, and C.~Patrignani are gratefully acknowledged.

% ****************************************************************************
% BIBLIOGRAPHY AREA
% ****************************************************************************

\begin{footnotesize}
% IF YOU DO NOT USE BIBTEX, USE THE FOLLOWING SAMPLE SCHEME FOR THE REFERENCES
% ----------------------------------------------------------------------------

% ----------------------------------------------------------------------------
% IF YOU USE BIBTEX,
% - DELETE THE TEXT BETWEEN THE TWO ABOVE DASHED LINES
% - UNCOMMENT THE NEXT TWO LINES AND REPLACE 'Name_Of_Your_BibFile'

%\bibliographystyle{unsrt}
%\bibliography{Name_Of_Your_BibFile}
% example of Name_Of_Your_BibFile.bib
% @Article{Turcato:2006ch,
%      author    = "Turcato, M.",
%  collaboration = "ZEUS and H1",
%      title     = "Lepton flavour violation and charmonium physics at HERA",
%      journal   = "Nucl. Phys. Proc. Suppl.",
%      volume    = "162",
%      year      = "2006", 
%      pages     = "283-287",
%      SLACcitation  = "%%CITATION = NUPHZ,162,283;%%"
% }
% 
% @Unpublished{Gogitidze:2007du,
%      author    = "Gogitidze, N.",
%  collaboration = "H1", 
%      title     = "Prompt photons and particle momentum distributions at
%                   HERA", 
%      year      = "2007",
%      note    = "hep-ex/0701033",
%      SLACcitation  = "%%CITATION = HEP-EX 0701033;%%"
% }

\end{footnotesize}

% ****************************************************************************
% END OF BIBLIOGRAPHY AREA
% ****************************************************************************

\end{document}